\documentclass[conference]{IEEEtran}
\usepackage{cite}
\usepackage{amsmath,amssymb,amsfonts}
\usepackage{algorithmic}
\usepackage{float}
\usepackage{graphicx}
\usepackage{textcomp}
\usepackage{xcolor}
\usepackage{algorithm}
\usepackage{multirow}
\usepackage{mathtools}
\def\BibTeX{{\rm B\kern-.05em{\sc i\kern-.025em b}\kern-.08em
    T\kern-.1667em\lower.7ex\hbox{E}\kern-.125emX}}
\DeclarePairedDelimiter{\ceil}{\lceil}{\rceil}

\begin{document}

\title{Enhancing Peak Network Traffic Prediction via Time-Series Decomposition
}

\author{\IEEEauthorblockN{Tucker Stewart}
\IEEEauthorblockA{\textit{School of Engineering and Technology} \\
\textit{University of Washington, Tacoma, WA} \\
trstew@uw.edu}
\and
\IEEEauthorblockN{Bin Yu}
\IEEEauthorblockA{\textit{Infoblox} \\
\textit{Santa Clara, CA}\\
biny@infoblox.com}
\and
\IEEEauthorblockN{Anderson Nascimento, Juhua Hu}
\IEEEauthorblockA{\textit{School of Engineering and Technology} \\
\textit{University of Washington, Tacoma, WA}\\
\{andclay, juhuah\}@uw.edu}
}

\maketitle

\begin{abstract}
For network administration and maintenance, it is critical to anticipate when networks will receive peak volumes of traffic so that adequate resources can be allocated to service requests made to servers. In the event that sufficient resources are not allocated to servers, they can become prone to failure and security breaches. On the contrary, we would waste a lot of resources if we always allocate the maximum amount of resources. Therefore, anticipating peak volumes in network traffic becomes an important problem. However, popular forecasting models such as Autoregressive Integrated Moving Average (ARIMA) forecast time-series data generally, thus lack in predicting peak volumes in these time-series. More than often, a time-series is a combination of different features, which may include but are not limited to 1) Trend, the general movement of the traffic volume, 2) Seasonality, the patterns repeated over some time periods (e.g. daily and monthly), and 3) Noise, the random changes in the data.  Considering that the fluctuation of seasonality can be harmful for trend and peak prediction, we propose to extract seasonalities to facilitate the peak volume predictions in the time domain. The experiments on both synthetic and real network traffic data demonstrate the effectiveness of the proposed method.
\end{abstract}

\section{Introduction}
\label{sec:introduction}
Many of previous works done for time-series forecasting have focused on general trend prediction~\cite{hoong:2012, stolojescu-crisan:2012, mozo:2018, purnawansyah:2018}. While these kind of forecasting models have been sufficient for most problems where entities are forecasting a target variable for any desirable point in time generally, these models lack in forecasting peak values in the data and often under predict these values. For some entities, this task is of even greater significance; namely network administrators of Domain Name System (DNS) and Dynamic Hosting Configuration Protocol (DHCP) servers.

For network administration and maintenance, it is critical to anticipate when networks will receive peak volumes of traffic so that adequate resources can be allocated to service requests made to servers.
Networks receive spikes in traffic for a number of reasons that are difficult to anticipate. Demand from customers for access to these servers and networks can see a rapid rise, thus traffic on the network increases. However, requests made to these servers can also be malicious in nature as these peaks in traffic volume can also be the consequences of malicious actors. One such way is a DNS Denial of Service (DDoS) attack. This is where a malicious actor hijacks the machines of benign users to create a bot net that then generates DNS queries towards the victim DNS server to overload it \cite{anagnostopoulos:2013, narfarieh:2019, truong:2016}. Despite the reason that network traffic receives these sudden increases in flow, the risks in not allocating enough resources to service these requests are the same. When resources are not sufficiently allocated to networks, they become prone to failure, render the server unresponsive to customers, and create a potential for security breaches. For this reason, the consequences of under predicting these peak values are much greater than over predicting.

However, popular forecasting models such as Autoregressive Integrated Moving Average (ARIMA) and Recurrent Neural Networks (RNN) have been focused on forecasting time-series data generally, thus lack in predicting peak volumes in the series \cite{hoong:2012, stolojescu-crisan:2012, mozo:2018, purnawansyah:2018, fu:2016, ergenc:2019, gracinai:2019}. ARIMA is a statistical model that forecasts values as a linear combination of previously observed values. Among many of ARIMA's strengths, it is great for modeling the trend and seasonality of a time-series \cite{chatfield:2019, shumway:2017, soyiri:2012} but lacks in forecasting the extrema within the data. This is due to network traffic being non-linear and also because peak values can be stochastic events that are not captured in the key systematic components of the data \cite{fu:2016}. %Perhaps by isolating each component of the time-series, such as the seasonalities, we can get better performance towards peak prediction.

Considering that the fluctuation of seasonality can be harmful for peak prediction, in this paper, we aim to study how time-series decomposition can be used to improve prediction when peak volumes occur in time-series. More than often, time-series are a combination of different features, which may include but are not limited to 1) Trend, the general movement of the traffic volume, 2) Seasonality, the patterns repeated over some time periods (e.g. daily and monthly), and 3) Noise, the random changes in the data \cite{pickering:2018}. Therefore, we propose to extract the seasonalities and study how forecasting these components independently in the time domain can be used to improve both the general time-series forecasting and the peak volume prediction.

To test the efficacy of time-series decomposition for general and peak prediction, we propose a framework that uses signal transformation techniques to decompose the time-series. Here, the Fourier transform is used to extract sinusoidal seasonal components from the trend and noise components. Then the trend plus noise component of the network traffic data are handled together and a model, such as ARIMA or Neural Networks, can be fitted to the data to forecast future observations in the time domain. We applied this decomposition framework to an Unresolved DNS Queries data set, our synthetic data set idealized for the Fourier Transform, and an electricity consumption data set as another real data set to evaluate our methodology. This empirical study demonstrates an increase in performance for peak prediction using our proposal.

\section{Related Work}

%%%%% Subsection: General Forecasting
\subsection{General Forecasting Models}

Most of the work done for modeling the flow of traffic interacting with a network has been focused on general trend prediction, that is, minimizing the error for all values rather than just peak values. Popular models that have been successful in modeling time-series can be divided into two categories: linear statistical models and non-linear models.

Many entities have had success in applying linear models such as ARIMA, Seasonal ARIMA (SARIMA), and other auto regressive models\cite{hoong:2012, purnawansyah:2018, ergenc:2019}. These models were one of the first to become popular because they are relatively easy statistical approaches to implement. Compared to other modeling techniques, they are relatively simple in their architecture and requires less data to train. While linear models have been successful in modeling network traffic effectively, due to their auto regressive nature in forecasting future observations as a linear combination of previous observations, they fail to represent the stochastic and non-linear dynamics of network traffic, namely the peak volumes of queries\cite{ergenc:2019}. For this reason, auto regressive models and other linear models are insufficient in capturing peak information.

To address the non-linear nature of network traffic, Neural Networks (NN) have been used as a suitable alternative to forecasting network traffic. In studies comparing ARIMA and variants of NN, NN has shown either marginal or considerable improvement to ARIMA for forecasting network traffic. There are, however, several different types of NNs that have been implemented in forecasting network traffic. The majority of architectures that have been applied to network traffic can be divided into two main families; Feed Forward Neural Networks containing ANNs~\cite{stolojescu-crisan:2012} and Convolutional Neural Networks (CNN)~\cite{mozo:2018}, and Recurrent Neural Networks such as Long-Short Term Memory (LSTM) and Gated Recurrent Units (GRU) \cite{fu:2016}.

In a study comparing the performance of ARIMA and ANN on WiMAX wireless network traffic, Stolojescu found that ANN was able to achieve better performance for forecasting small future time intervals compared to ARIMA \cite{stolojescu-crisan:2012}. RNNs may be preferable due to their ability to memorize information about previously observed data \cite{fu:2016}. This is important to capture for time-series forecasting due to the temporal dependencies present in time-series data. This has made RNN a prime candidate for forecasting network traffic. Fu et al.~\cite{fu:2016} was able to achieve slight decreases in the MSE and Mean Absolute Error (MAE) using LSTM and GRU over ARIMA. However, most of the existing work has been focusing on the general prediction rather than peak volume prediction of this work.%RNNs were not used in this empirical study as they demand vast quantities of data to be effective that not all data sets, namely the Unresolved DNS Queries data set, could provide.

%%%%% Subsection: Time-Series Decomposition
\subsection{Time-Series Decomposition}

Time-series Decomposition bases itself on the concept that time-series are comprised of several different components either additively or multiplicatively. The most prominently identified components in time-series decomposition include, but are not limited to, the Trend, Seasonality, and the Residual. The Trend is the general movement the series follows. Over time, does the data tend to increase or decrease overall and at what rate? Next the data can have multiple seasonal components which are the patterns that repeat periodically for some time interval; namely daily, weekly, monthly, annually, or following with the seasons. Lastly, there is the residual, also known as the noise or error within the data. These are the random changes that cause the data to deviate from the recognizable patterns. Decomposition is the process in which we separate the time-series into smaller components that contribute to the overall result \cite{pickering:2018}. The motivation behind doing so is that by decomposing the series into individual pieces, they can then be isolated and have a model fitted to identify that particular pattern to improve performance of the overall forecasting model\cite{purnawansyah:2018}.

One method for decomposing a time-series is to apply signal processing transformations to the series to transform the series into another base. Then, the information presented in the new base can be used to extract the components from the original series. Fourier Transform is one of these methods. Concretely, the Fourier Transform can be used to transform a time-series from the time domain to the frequency domain. The series is then represented as the sum of sinusoids (i.e., sine and cosine waves). This is particularly useful for identifying and extracting seasonal components from the times series since the sine and cosine waves are patterns that repeat periodically, which fits our definition of seasonal components. Lewis et al.~\cite{lewis:2003} found that they were able to accurately forecast the volume of call received by a call centre by applying the Fourier Transform and forecasting in the frequency domain rather than the time domain. However, not much work has been done to apply the Fourier Transform on network traffic or specifically for the problem of peak volume prediction. 

It should be emphasized that Lewis et al.~\cite{lewis:2003} also applied the Fourier Transform to their call centre data to compute the corresponding frequency spectrum which was inputted directly into the forecasting function $F$ as input. In this work, the Fourier Transform is merely used as a decomposition technique for separating seasonal components from the data. The components are modeled and forecast in the time domain rather than the frequency domain. Additionally, the seasonal components are calculated mathematically rather than being forecast by a model.
%%%%%%%%%%%%%%%%%%%%%%%%%%%%%%%%%%%%%%%%%%%%%%%%%%%%%%%%%%%%%%%%%%%%%%%%%%%%%
%%%%% Section: The Proposed Method
\section{The Proposed Method}

Given a univariate time-series of the form
\[X =\{ \cdots, x_{t-1}, x_{t}, x_{t+1}, \cdots\}\]
where $x_{t}$ is the quantity of the measured variable (e.g., the traffic volume) at time $t$. The process of forecasting the traffic volume of a network at time $t+1$ will receive can be denoted as
% Equation: Forecasting
\begin{equation}
    x_{t+1} = F(x_{t}, x_{t-1}, \cdots, x_{t-w+1})
    \label{eq:forecasting}
\end{equation}
where $F$ is a general time-series forecasting algorithm and $w$ is the window size denoting the number of past observations used for forecasting.

Instead of learning $F$ directly from the observed time-series, we apply the Fast Fourier Transform \cite{christensen:2010} to transform $X$ from the time domain into the frequency domain. In the frequency domain, the time-series is represented as the sum of complex sinusoids, that is, sine and cosine waves. The high amplitude sinusoids are likely seasonal components with a pattern repeating within a regular period~\cite{lewis:2003, jin:2015}. Therefore, we propose to extract only the first $c$ highest amplitude sinusoids excluding the zero frequency sinusoid. These sinusoids with a given frequency, amplitude, and phase, can easily be calculated as cosines for the desired time interval to forecast.

Then, the remaining components can be transformed back into the time domain as
\[X' =\{ \cdots, x_{t-1}', x_{t}', x_{t+1}', \cdots\}\] 
For them, we aim to learn the prediction function as
\begin{equation}
    x_{t+1}' = F(x_{t}', x_{t-1}', \cdots, x_{t-w+1}')
    \label{eq:forecasting1}
\end{equation}
Before fitting the model, we prepare the data $X'$ using scaling and normalization. Inspired by a previous work~\cite{gracinai:2019}, we apply local normalization to better capture the peak information. However, before the local normalization can be applied, the time-series $X'$ needs to be scaled to [1, 2] using Eqn.~\ref{eq:minmax_scaling}.
% Equation: MinMax Scaling
\begin{equation}
    s_{t} = \frac{x_{t}' - X_{min}'}{X_{max}' - X_{min}'} + 1
    \label{eq:minmax_scaling}
\end{equation}

This scaling step is important considering the special property of local normalization in Eqn.~\ref{eq:local_normalization}. Specifically, we need to ensure that the data does not have any values between $[0, 1)$ since local normalization divides the current observation by its previous observation and values in $[0, 1)$ can greatly distort the scale as dividing a value by a number between zero and one is the same as multiplying the value by a number greater than one. While local normalization tends to perform greater in terms of peak prediction, it can also suppress seasonality that exists within the time-series. Fortunately, the seasonal components have been removed and will be forecast independently in our proposed methodology.

For local normalization, each data point is normalized by the previously observed value, such that the normalized time-series,
\[L = \{\cdots, l_{t-1}, l_{t}, l_{t+1}, \cdots\}\] 
is calculated by
% Equation: Local Normalization
\begin{equation}
    l_{t} = \frac{s_{t} - s_{t-1}}{s_{t-1}}
    \label{eq:local_normalization}
\end{equation}
Finally, the data $L$ is scaled into the range of [-1, 1] using Eqn.~\ref{eq:maxabs_scaling} to ensure that the scale of the values does not affect the model estimation.
% Equation: Absolute Max Scaling
\begin{equation}
    y_{t} = \frac{l_{t}}{|L_{max}|}
    \label{eq:maxabs_scaling}
\end{equation}

\begin{algorithm}[t]
   \caption{PPFD}
   \label{alg: ppfd}
\begin{algorithmic}
   \STATE {\bfseries Input:} Training Time Series $X =\{ x_{t1}, x_{t2},\cdots, x_{tn}\}$, the number of seasonal components $c$.
   \STATE Apply FFT to $X$
   \STATE Filter out the first $c$ highest amplitude components
   \FOR{$season=1$ {\bfseries to} $c$}
   \STATE Compute the cosine wave
   \ENDFOR
   \STATE Run inverse FFT on the remaining frequency spectrum to get the time series $X'$ without the seasonal components
   \STATE Scale $x_t'$ to $s_t\in [1, 2]$ by Eqn.~\ref{eq:minmax_scaling}
   \STATE Local normalization $s_t$ to $l_t$ by Eqn.~\ref{eq:local_normalization}
   \STATE Re-scale $l_t$ to $y_t\in [-1, 1]$ by Eqn.~\ref{eq:maxabs_scaling}
   \STATE Fit a forecasting model $F$ based on $y_t$'s
   \RETURN $c$ cosine waves and model $F$
\end{algorithmic}
\end{algorithm}

In summary, to forecast the future volumes of $x_{t+1}'$ in Eqn.~\ref{eq:forecasting1}, we will learn a prediction model as 
\begin{equation}
    y_{t+1} = F(y_{t}, y_{t-1}, \cdots, y_{t-w+1})
    \label{eq:forecasting2}
\end{equation}
Our proposed peak prediction framework, i.e., Peak Prediction via Fourier Decomposition (PPFD), is summarized in Alg.~\ref{alg: ppfd}. Now given an observed time series, we can predict the upcoming volume by summing over the predictions from each cosine wave and that $y_{t+1}$ from the model $F$. It should be notes that $y_{t+1}$ should be firstly de-scaled and de-normalized in the opposite order to retrieve the forecast $x_{t+1}'$.

%In the following subsection, we illustrate the main procedure of the proposed method using the synthetic data.

%%%%%%%%%%%%%%%%%%%%%%%%%%%%%%%%%%%%%%%%%%%%%%%%%%%%%%%%%%%%%%%%%%%%%%%%%%%%%
% Figures and Statistics for the data sets of the Experiment Section.
\begin{table*}[!ht]
    \centering
    \caption{Statistics of the Data Sets}
    \label{tab:data_stats}
    \begin{tabular}{|l|c|c|c|c|c|}
        \hline
        \textbf{Data Set} & \textbf{Data Points} & \textbf{Mean} & \textbf{Min} & \textbf{Median} & \textbf{Max} \\
        \hline
        Synthetic & 7500 & 1375816074.4 & 824704284.8 & 1375148204.1 & 1944188859.3 \\
        \hline
        Unresolved DNS Queries & 968 & 105400760.9 & 46 & 98374184 & 294870823 \\
        \hline
        Unresolved DNS Queries (Linear Interpolation) & 1040 & 100498446.9 & 46 & 94268425 & 294870823 \\
        \hline
        % Carriageway Traffic & 34321 & 568.9 & 19.0 & 624.0 & 1307.5 \\
        % \hline
        % Carriageway Traffic (Subset) & 6285 & 600.4 & 38.5 & 673.8 & 1286.3 \\
        % \hline
        Electricity Consumption & 32588 & 1.07789 & 0.0 & 0.78252 & 6.56053 \\
        \hline
        Electricity Consumption (Linear Interpolation) & 32164 & 1.09039 & 0.0296 & 0.79092 & 6.56053 \\
        \hline
        % Synthetic, Trend & 7500 & 1374950000 & 1000000000 & 1374950000 & 1749900000 \\
        % \hline
        % Synthetic, Trend with Weekly Seasonality & 7500 & 1374968673.9 & 922505846.9 & 1374854662.0 & 1828012289.8 \\
        % \hline
    \end{tabular}
\end{table*}

\begin{figure*}[!ht]
    \centering
    \includegraphics[width=0.97\textwidth]{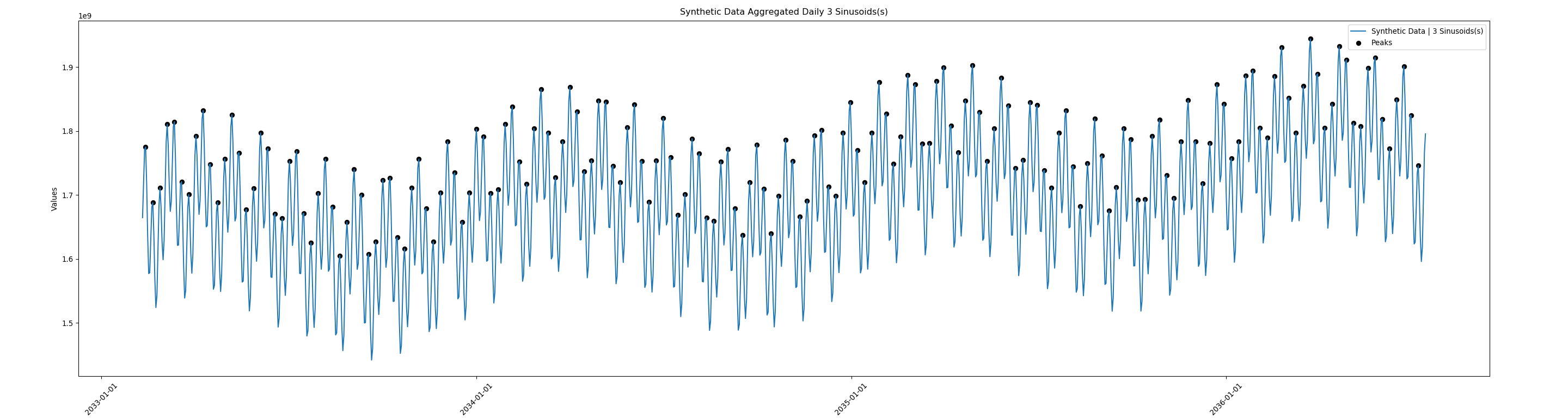}
    \caption{Synthetic Data Set Aggregated Daily. (Last fold from the cross validation)}
    \label{fig:synthetic_plot3}
\end{figure*}

\begin{figure}[!ht]
    \centering
    \includegraphics[width=0.48\textwidth]{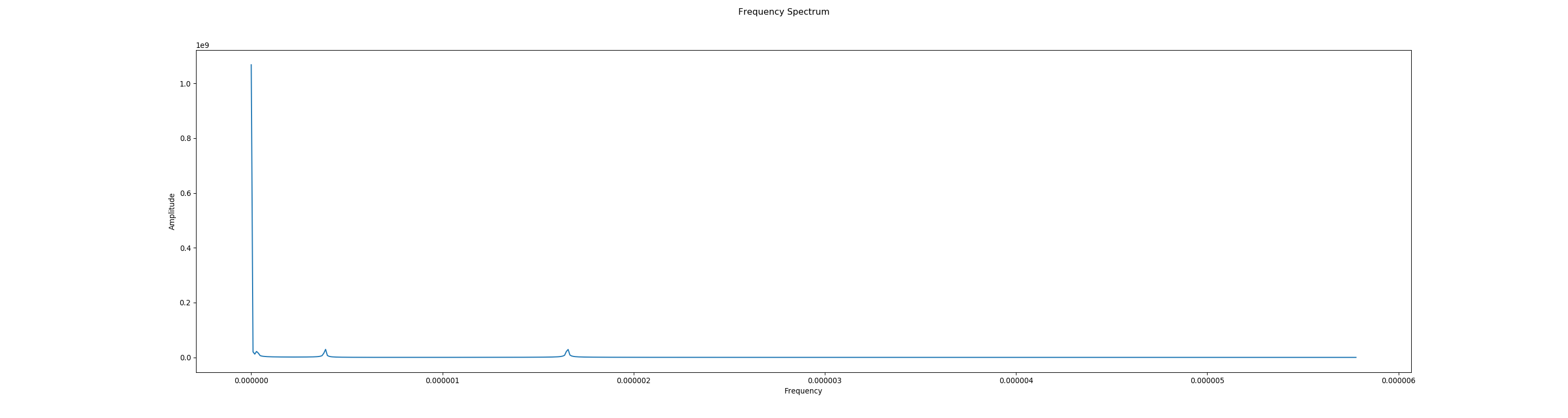}
    \caption{Frequency domain of the synthetic data}
    \label{fig:synthetic_freq}
\end{figure}
\begin{figure}[!ht]
    \centering
    \includegraphics[width=0.48\textwidth]{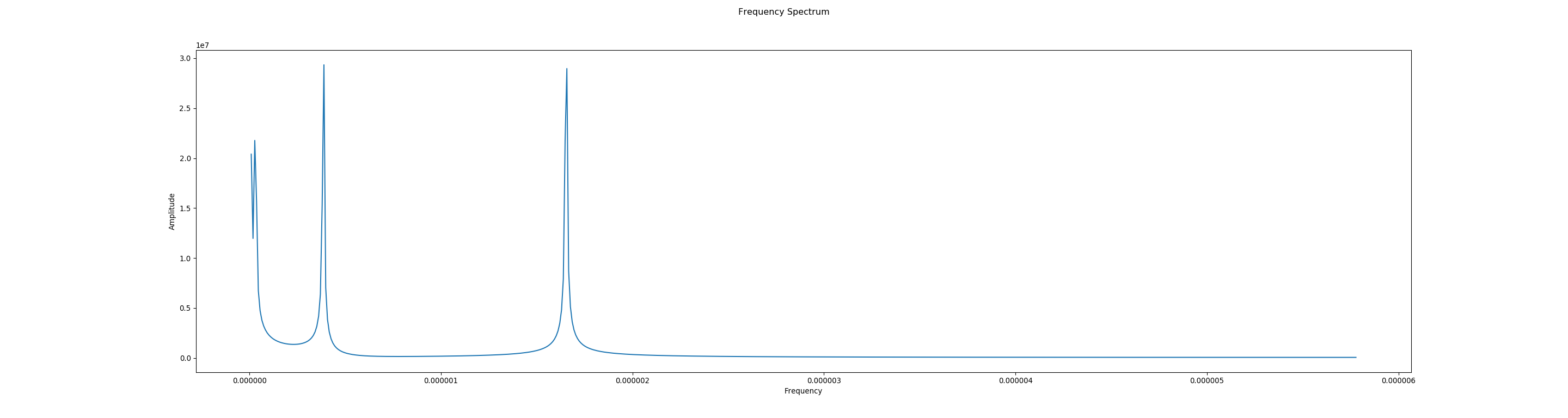}
    \caption{Frequency domain, with the zero frequency component removed, of the synthetic data.}
    \label{fig:synthetic_freq_zoom}
\end{figure}
\begin{figure}[!ht]
    \centering
    \includegraphics[width=0.48\textwidth]{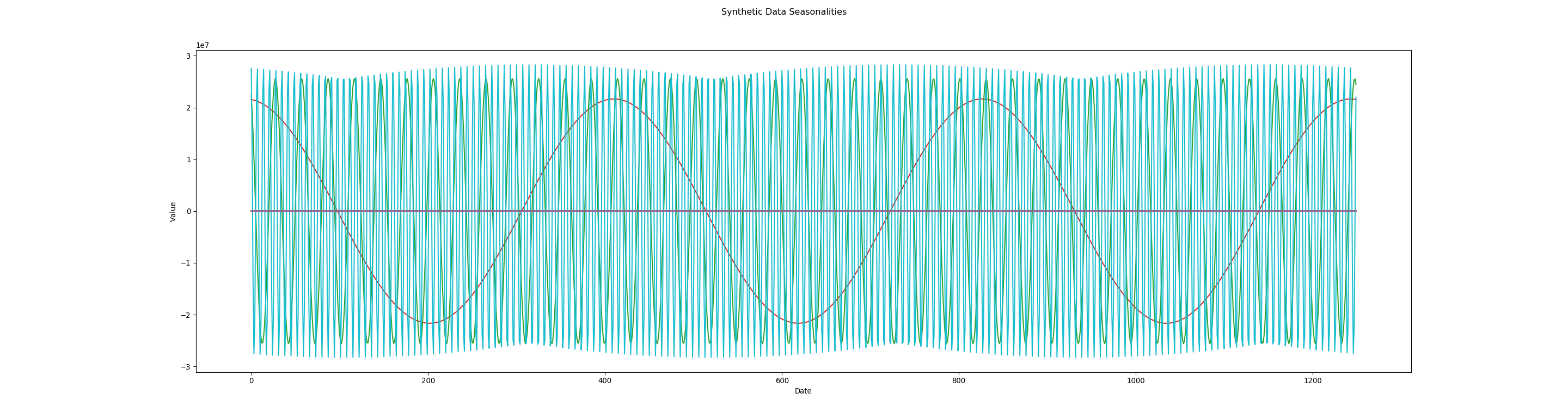}
    \caption{The three sinusoids extracted from the synthetic data set calculated as cosine waves.}
    \label{fig:synethic_sinusoids}
\end{figure}
\begin{figure}[!ht]
    \centering
    \includegraphics[width=0.48\textwidth]{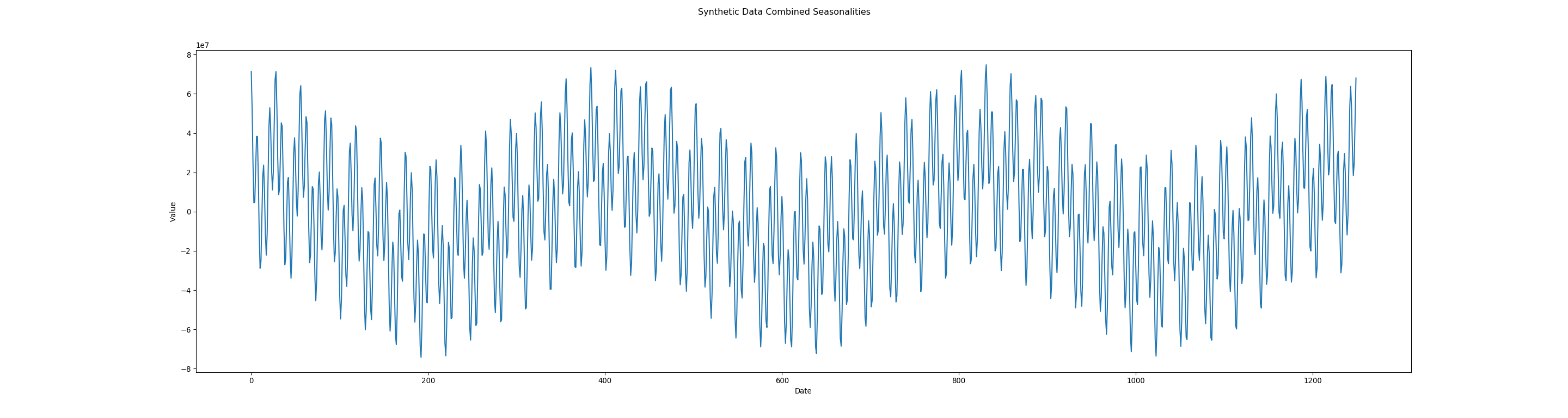}
    \caption{Sum of the three extracted sinusoids from the synthetic data}
    \label{fig:synethic_combined_sinusoids}
\end{figure}
\begin{figure}[!ht]
    \centering
    \includegraphics[width=0.48\textwidth]{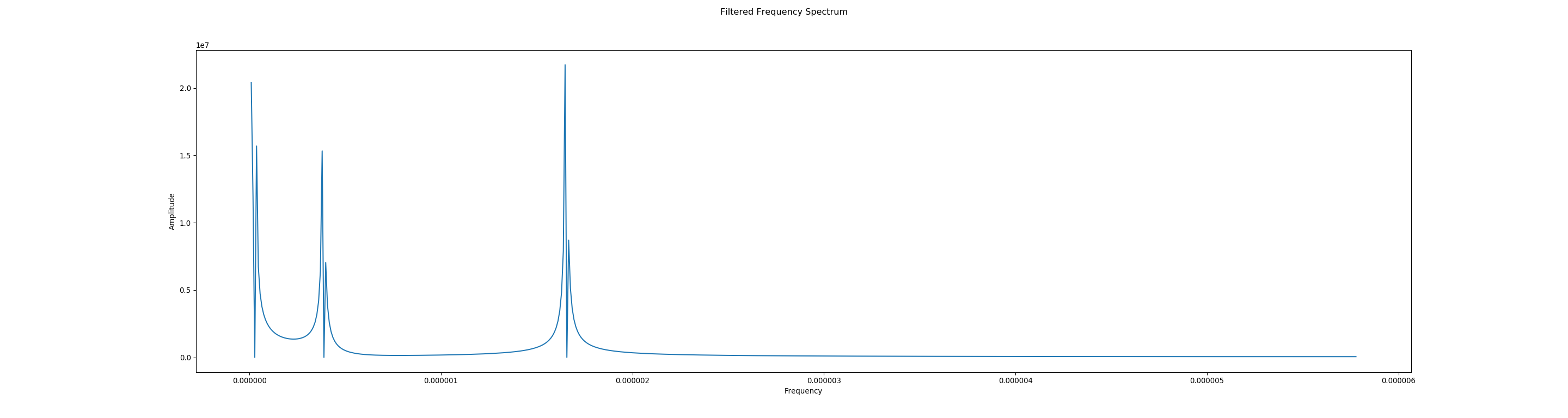}
    \caption{Frequency domain of the synthetic data with the three highest amplitude sinusoids removed}
    \label{fig:synthetic_freq_filtered}
\end{figure}

\begin{table*}[!ht]
    \centering
    \caption{Performance Comparison on Synthetic Data}
    \label{tab:synthetic_results3}
    \begin{tabular}{|c|c|c|c|c|c|c|c|}
        \hline
        \textbf{Model} & \textbf{c} & \textbf{RMSE} & \textbf{RWSE} & \textbf{Peak RMSE} & \textbf{Peak RWSE} & \textbf{Under Predicted} & \textbf{Over Predicted} \\
        \hline
        \multicolumn{8}{c}{\textbf{Baseline Models}} \\
        \hline
        ANN &  & 0.02442 & 0.01908 & 0.01007 & 0.00994 & 630 & 261 \\
        \hline
        ARIMA &  & 0.02395 & 0.01874 & 0.00913 & 0.00913 & 883 & 8 \\
        \hline
        FOURIER &  & 0.03366 & 0.02597 & 0.01276 & 0.01220 & 723 & 168 \\
        \hline
        \multicolumn{8}{c}{\textbf{Fourier Decomposition}} \\
        \hline
        \multirow{3}{*}{PPFD with ANN} & 1 & 0.02242 & 0.01755 & 0.00958 & 0.00944 & 639 & 252 \\
        & 2 & \textbf{0.02226} & \textbf{0.01741} & 0.00921 & 0.00853 & 561 & 330 \\
        & 3 & 0.02397 & 0.01877 & \textbf{0.00873} & \textbf{0.00659} & \textbf{447} & \textbf{444} \\
        \hline
        \multirow{3}{*}{PPFD with ARIMA} & 1 & 0.02961 & 0.02091 & 0.01805 & 0.01314 & 782 & 109 \\
        & 2 & 0.02873 & 0.01975 & 0.01937 & 0.01266 & 654 & 237 \\
        & 3 & 0.02992 & 0.02014 & 0.01940 & 0.01513 & 449 & 442 \\
        \hline
    \end{tabular}
\end{table*}

\begin{figure*}[ht]
    \centering
    \includegraphics[width=\textwidth]{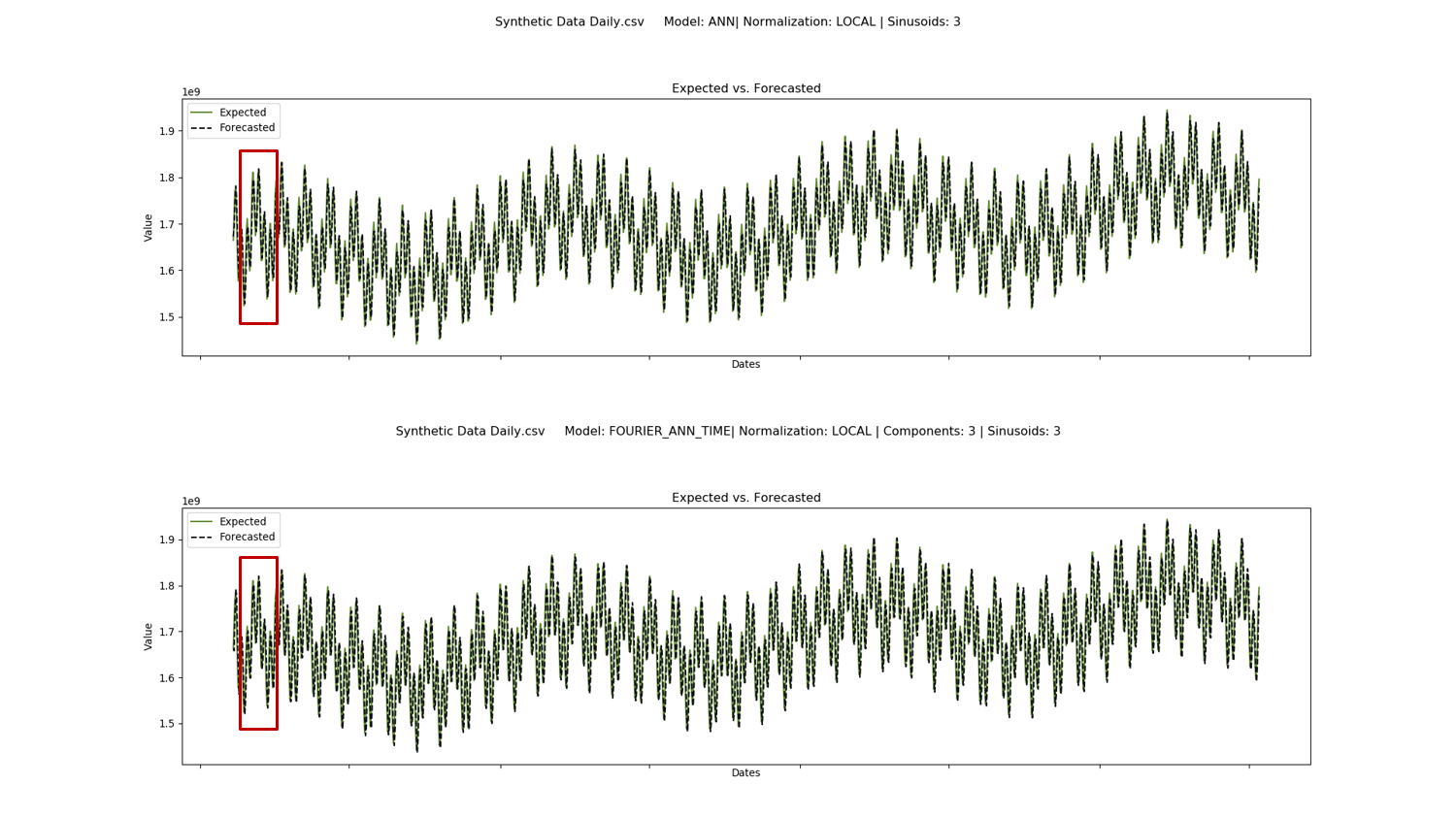}
    \caption{Expected vs Forecast: ANN (Top) vs. PPFD with ANN $(c = 3)$ (Bottom) on Synthetic Data.}
    \label{fig:syn_ann_vs_fourier_ann}
\end{figure*}

\begin{figure*}[ht]
    \centering
    \includegraphics[width=\textwidth]{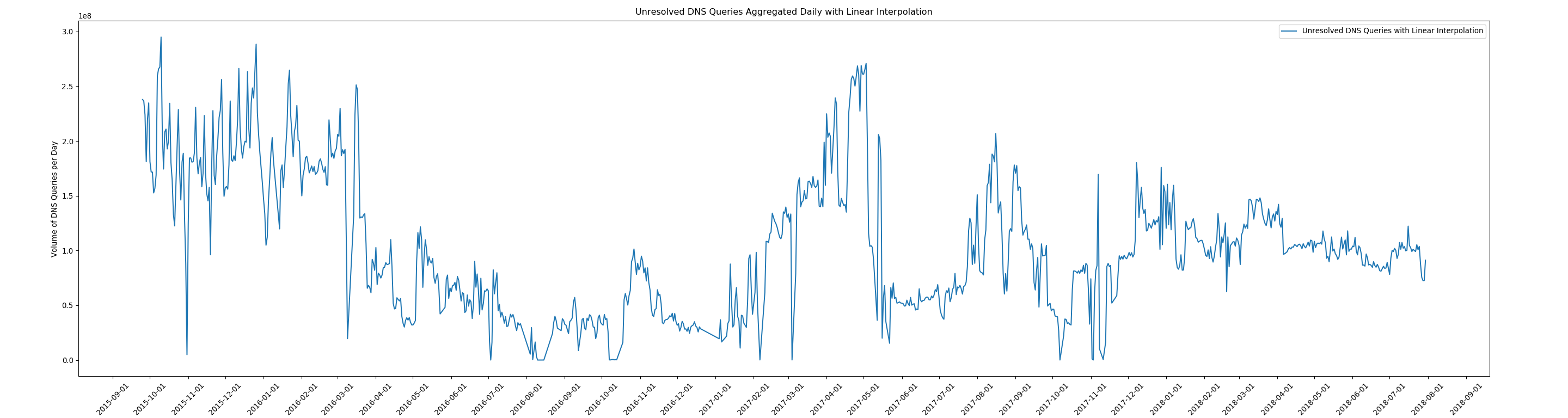}
    \caption{Daily Unresolved DNS Queries after applying Linear Interpolation}
    \label{fig:dns_linear_plot}
\end{figure*}

% Results Table: Unresolved DNS Queries with Linear Interpolation
\begin{table*}[ht]
    \centering
    \caption{Performance Comparison on Unresolved DNS Queries with Linear Interpolation}
    \label{tab:dns_results_linear}
    \begin{tabular}{|c|c|c|c|c|c|c|c|}
        \hline
        \textbf{Model} & \textbf{c} & \textbf{RMSE} & \textbf{RWSE} & \textbf{Peak RMSE} & \textbf{Peak RWSE} & \textbf{Under Predicted} & \textbf{Over Predicted} \\
        \hline
        \multicolumn{8}{c}{\textbf{Baseline Models}} \\
        \hline
        ANN &  & 0.17796 & \textbf{0.13408} & 0.23630 & 0.23602 & 105 & 5 \\
        \hline
        ARIMA &  & 0.18932 & 0.13761 & 0.23634 & 0.23588 & 101 & 9 \\
        \hline
        FOURIER &  & 0.18923 & 0.13790 & 0.23430 & 0.23422 & 105 & 5 \\
        \hline
        \multicolumn{8}{c}{\textbf{Fourier Decomposition}} \\
        \hline
        \multirow{4}{*}{PPFD with ANN} & 3 & 0.18004 & 0.13416 & 0.23481 & 0.23429 & 105 & 5 \\
        & 5 & 0.17896 & 0.13463 & 0.23344 & 0.23330 & 102 & 8 \\
        & 7 & \textbf{0.17649} & 0.13433 & 0.23385 & 0.23376 & 100 & 10 \\
        & 10 & 0.18272 & 0.13757 & 0.24157 & 0.24150 & 105 & 5 \\
        % & 40 & 0.19553 & 0.14944 & 0.24987 & 0.24957 & 102 & 8 \\
        \hline
        \multirow{4}{*}{PPFD with ARIMA} & 3 & 0.19745 & 0.13973 & 0.23294 & 0.23160 & 96 & 14 \\
        & 5 & 0.19616 & 0.13978 & 0.23360 & 0.23238 & 97 & 13 \\
        & 7 & 0.19636 & 0.14014 & 0.23353 & 0.23218 & 96 & 14 \\
        & 10 & 0.19593 & 0.13999 & \textbf{0.23176} & \textbf{0.23080} & \textbf{96} & \textbf{14} \\
        % & 40 & 0.20708 & 0.14863 & 0.24413 & 0.24318 & 94 & 16 \\
        \hline
    \end{tabular}
\end{table*}

\begin{figure*}[ht]
    \centering
    \includegraphics[width=\textwidth]{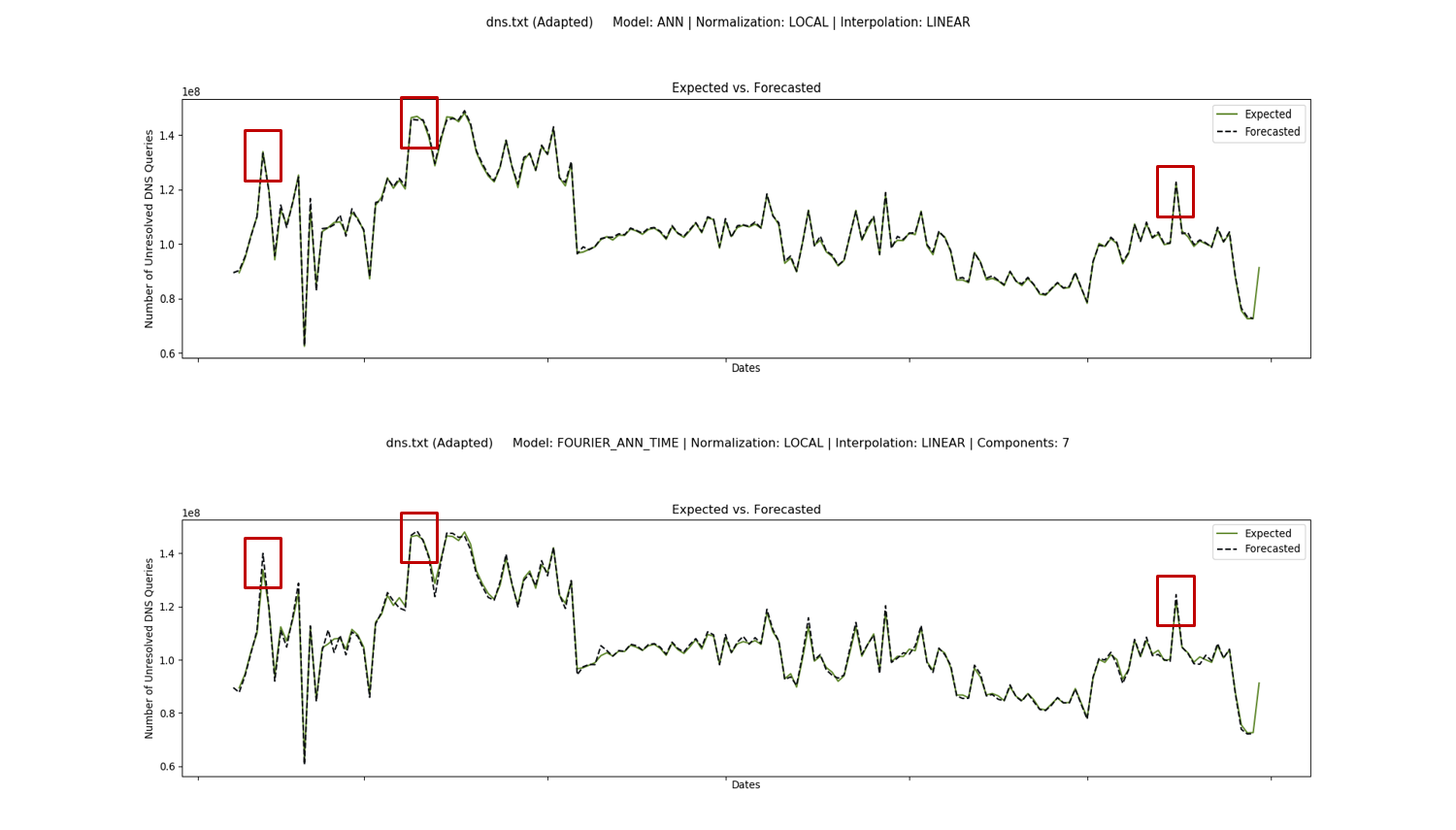}
    \caption{Expected vs Forecasted: ANN (Top) vs. PPFD with ANN $(c = 7)$ (Bottom) on Unresolved DNS Queries. (The time lag of one day was reduced for the sake of this visualization only.)}
    \label{fig:dns_ann_vs_fourier_ann}
\end{figure*}

\begin{figure*}[ht]
    \centering
    \includegraphics[width=\textwidth]{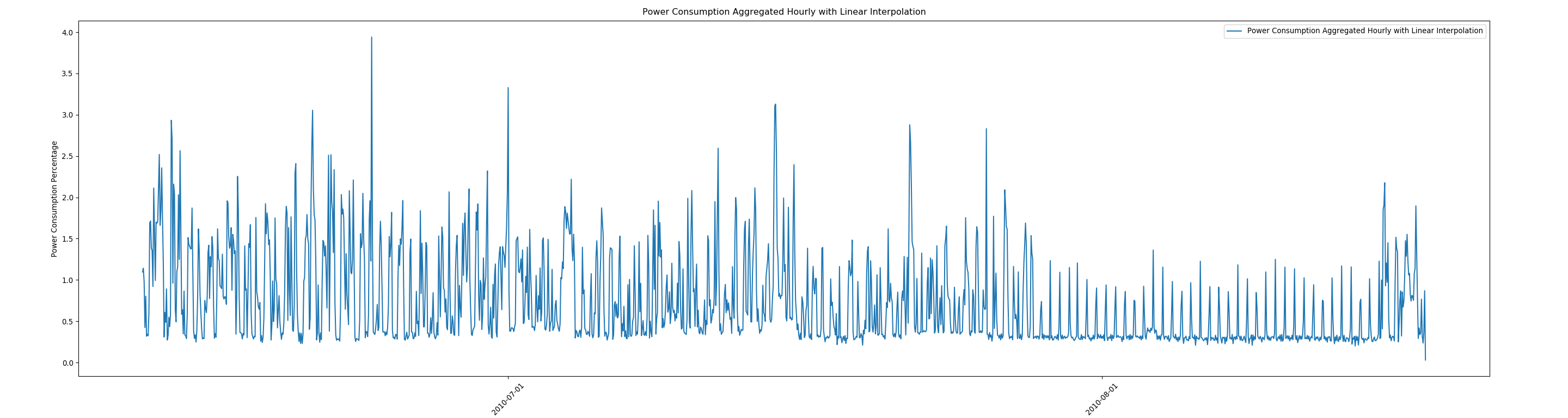}
    \caption{Hourly electricity consumption data after applying Linear Interpolation}
    \label{fig:electricity_linear_plot}
\end{figure*}

% Results Table: Electricity Consumption with Linear Interpolation
\begin{table*}[ht]
    \centering
    \caption{Performance Comparison on Electricity Consumption Data with Linear Interpolation}
    \label{tab:hourly_power_linear}
    \begin{tabular}{|c|c|c|c|c|c|c|c|}
        \hline
        \textbf{Model} & \textbf{c} & \textbf{RMSE} & \textbf{RWSE} & \textbf{Peak RMSE} & \textbf{Peak RWSE} & \textbf{Under Predicted} & \textbf{Over Predicted} \\
        \hline
        \multicolumn{8}{c}{\textbf{Baseline Models}} \\
        \hline
        ANN &  & 0.61432 & 0.48528 & 0.88223 & 0.88217 & 3599 & 120 \\
        \hline
        ARIMA &  & 0.66396 & 0.49462 & 0.86497 & 0.86390 & 3508 & 211 \\
        \hline
        FOURIER &  & 0.62570 & 0.48862 & 0.88123 & 0.88113 & 3556 & 163 \\
        \hline
        \multicolumn{8}{c}{\textbf{Fourier Decomposition}} \\
        \hline
        \multirow{4}{*}{PPFD with ANN} & 3 & \textbf{0.61333} & \textbf{0.48338} & 0.87782 & 0.87776 & 3523 & 196 \\
        & 5 & 0.61652 & 0.48544 & 0.87876 & 0.87868 & 3510 & 209 \\
        & 7 & 0.61495 & 0.48621 & 0.88115 & 0.88104 & 3496 & 223 \\
        & 10 & 0.61846 & 0.48619 & 0.87673 & 0.87662 & 3503 & 216 \\
        % & 40 & 0.62488 & 0.49145 & 0.88062 & 0.88249 & 3485 & 234 \\
        \hline
        \multirow{4}{*}{PPFD with ARIMA} & 3 & 0.65982 & 0.49305 & 0.86716 & 0.86639 & 3480 & 239 \\
        & 5 & 0.66156 & 0.49391 & 0.86633 & 0.86543 & 3481 & 238 \\
        & 7 & 0.66307 & 0.49481 & \textbf{0.86484} & \textbf{0.86377} & \textbf{3457} & \textbf{262} \\
        & 10 & 0.66459 & 0.49573 & 0.86622 & 0.86511 & 3457 & 262 \\
        % & 40 & 0.67072 & 0.49959 & 0.86735 & 0.86593 & 3436 & 283 \\
        \hline
    \end{tabular}
\end{table*}
%%%%%%%%%%%%%%%%%%%%%%%%%%%%%%%%%%%%%%%%%%%%%%%%%%%%%%%%%%%%%%%%%%%%%%%%%%%%%
%%%%% Section: Experiment
\section{Experiment}\label{sec:exp}
To demonstrate the effectiveness of our proposed forecasting method, we evaluate it on both synthetic and real data sets. Each of these data sets is a univariate time-series aggregated at different time intervals. The statistics of each data set are summarized in Table \ref{tab:data_stats}.

%%%%% Subsection: Setup
\subsection{Setup}

As a baseline for the results of our proposed methodology, we apply ARIMA, ANN, and Fourier Forecasting independently from ARIMA and ANN as follows.

% List: Baseline Models
\begin{itemize}
    \item ANN~\cite{stolojescu-crisan:2012}
    \item ARIMA~\cite{chatfield:2019} with the optimal parameters selected.
    \item Fourier Forecasting where all sinusoids are forecasted and summed together.
\end{itemize}

In comparison to these baseline models, we will see the effectiveness our proposed framework through the following experiments.

% List: Proposed Models
\begin{itemize}
    \item PPFD using ANN~\cite{stolojescu-crisan:2012} to forecast the non-seasonal components.
    \item PPFD using ARIMA~\cite{chatfield:2019} to forecast the non-seasonal components.
\end{itemize}

For all of these experiments, only two general time-series forecasting models are used, that is, ANN and ARIMA. These models are setup for each experiment as follows.

\begin{itemize}
     \item ANN: The architecture of the model consists of a single input layer, one hidden layer made up of five sigmoid units, and an output layer with a single linear unit. The size of the input layer will vary depending on what kind of seasonality we expect the data to exhibit. For the Unresolved DNS queries and our synthetic data set, the input layer will have seven units to capture the weekly seasonality.
    \item ARIMA: ARIMA requires the parameters (P, D, Q). All data sets required one level of differencing so D is always set to 1. The number of parameters, P and Q, vary for the data set so the best parameters are selected for each data set \cite{chatfield:2019}.
\end{itemize}

%%%%% Subsection: Evaluation Metrics
\subsection{Evaluation Metrics}

RMSE is the standard evaluation metric for regression problems such as time-series forecasting. While this evaluation metric is sufficient for general prediction performance, it weights under prediction equally to over prediction and thus is not well suited for peak prediction. Concretely, the Mean Squared Error calculates the mean of squared errors as
% Equation: Mean Squared Error
\begin{equation}
    MSE = \frac{1}{N} \sum^{N}_{i = 1}{(\hat{x}_i-x_i)^2}
    \label{eq:mse}
\end{equation}
where $x_i$ is observed value and $\hat{x}_i$ is the forecasted value provided by the model.

Bin et al.~\cite{gracinai:2019} proposed a new cost-adaptive loss function that weights under prediction greater than over prediction called the Weighted Sign Error (WSE). Then, to penalize under prediction more heavily than over prediction, the loss function is adapted to consider the sign of the error for each observation, resulting in the following function as
% Equation: Weighted Sign Error
\begin{equation}
    WSE = \frac{1}{N} \sum^{N}_{i = 1}{\alpha^{\frac{1+sign(\hat{x}_i-x_i)}{2}} (\hat{x}_i-x_i)^2}
    \label{eq:wse}
\end{equation}
$\alpha \in [0, 1]$ is a weighted coefficient that determines the weight of over prediction. For our experiments, $\alpha$ is set to $0.2$, and $sign()$ is a function that returns a numerical value based on the sign of the error. Therefore, $sign(\hat{x}_i-x_i)$ returns $1$ if $\hat{x}_i \geq x_i$ and $-1$ otherwise.

To evaluate the performance of each model, we use Root Mean Squared Error (RMSE) and Root Weighted Sign Error (RWSE) defined as follows.
% Equation: Root Mean Squared Error
\begin{equation}
    RMSE = \sqrt{MSE}
    \label{eq:rmse}
\end{equation}
% Equation: Root Weighted Squared Error
\begin{equation}
    RWSE = \sqrt{WSE}
    \label{eq:rwse}
\end{equation}

We apply both of these evaluation methods on the entire validation set to evaluate both the general prediction and peak prediction as in~\cite{gracinai:2019}. In addition to these evaluation metrics, we use a statistical method to identify the peaks in the time-series (i.e., \textit{Find Peaks} function in Scipy's Signal module) and track the number of values in the validation data set that are under predicted and over predicted, whose RMSE and RWSE are also separately calculated to better capture the performance on peak predictions.

To test the efficacy of each model, we need to use some form of cross validation. Since time-series data are ordered and there is essential information in the time lags between data points, shuffling the data for k-fold cross validation is not desirable. We used time-series cross validation, also known as forward chaining cross validation~\cite{hermias:2017}. The number of rounds used in our experiments is five. All evaluation values are aggregated across all five folds and the average of each error, across the five folds is recorded.
%%%%%%%%%%%%%%%%%%%%%%%%%%%%%%%%%%%%%%%%%%%%%%%%%%%%%%%%%%%%%%%%%%%%%%%%%%%%%
%%%%% Section: Data Description
%\newpage
\subsection{Data Description and Performance Comparison}

%%%%% Subsection: Synthetic Data Set
\subsubsection{Synthetic Data Set}

We generated a synthetic data set free of noise, to test our proposed PPFD framework on data that has very clear sinusoidal seasonal components. This data is generated by calculating a linear trend and adding sine waves of a given amplitude and periodicity. To match with the Unresolved DNS query data set, this data set is sampled daily. The trend is calculated by the function $y_t = m*x_t + b$ where the slope, $m$, is 100,000 and the y-intercept, $b$, is 1,000,000,000. Then for the seasonality, we added three seasonal components. Since the Fourier Transform represents the frequencies as complex sinusoids, we generated the seasonal components as sine waves for the periods of weekly, monthly, and yearly and with amplitudes of 80,000,000, 72,000,000, and 56,000,000 respectively. These components are then added together to get the resulting synthetic data set depicted in Fig.~\ref{fig:synthetic_plot3}, where the dots denote the peaks found by the \textit{Find Peaks} function.% To test the effects of seasonality being present, we've applied our framework on the data set for trend with weekly seasonality and trend with all seasonal components.

On this synthetic data, we demonstrate step by step to illustrate the procedure of the proposed PPFD framework as follows.

%%%%% Subsection: Fourier Forecasting
%\subsection{Peak Prediction via Fourier Decomposition (PPFD)}

% List: Fourier Forecasting Procedures.
\begin{enumerate}
    \item After applying the Fast Fourier Transform (FFT) to the training time-series $X$, we obtain the frequency spectrum as in Fig.~\ref{fig:synthetic_freq}. It should be noted that the zero frequency component is not a seasonal component and the seasonalities can be emphasized by removing the zero component as in Fig.~\ref{fig:synthetic_freq_zoom}, where the three seasonal components manually added can be easily observed. 
    \item Then, we can filter out the first $c=3$ highest amplitude components, each of which will be used to forecast the corresponding seasonal component as follows. 
    \begin{enumerate}
        \item For each of the seasonal component, compute the amplitude from the complex numbers of the frequency spectrum and the phase shift.
        \item With the amplitude, frequency, and phase shift, extrapolate each seasonal component as a cosine wave over the forecasting time interval as in Fig.~\ref{fig:synethic_sinusoids}.
        \item Sum the seasonal components together to get the combined seasonality as in Fig.~\ref{fig:synethic_combined_sinusoids}.
    \end{enumerate}%each of the individual seasonal components that were extracted from the frequency spectrum.. 
    \item Set the $c=3$ highest amplitude components to zero in the frequency domain as done in Fig.~\ref{fig:synthetic_freq_filtered}. Run inverse FFT on the filtered frequency spectrum to get the time-series $X'$ in the time domain with the seasonal components removed, which will be scaled and normalized to train model $F$ to do predictions on this component separately.
\end{enumerate}

Table~\ref{tab:synthetic_results3} shows the prediction results of different methods, where $c$ indicates the number of highest seasonal components extracted. For the baseline FOURIER model, all the sinusoids are forecasted giving us a $c$ of $\ceil{\frac{N}{2}}$, where $N$ is the number of observations in the time-series. First, it can be easily observed that PPFD with ANN can not only improve forecasting performance of the peak volumes but also that of the general prediction, which demonstrates that seasonal fluctuations can be harmful not just for the peak predictions but also for the general predictions. Second, with the help of Fourier decomposition where the exact number (i.e., $c=3$) of seasonal components are extracted, the performance on peaks only (i.e., Peak RMSE, Peak RWSE) is significantly improved and the number of under predicted peaks is significantly decreased. This can also be observed in Fig.~\ref{fig:syn_ann_vs_fourier_ann}, where one example in red rectangular of peak volume that is under predicted by the baseline of ANN, is well predicted by PPFD with ANN. Therefore, for time-series with clear seasonality and no noise, forecasting seasonality and trend separately can help in both general prediction and peak prediction.

%%%%% Subsection: Unresolved DNS Query Traffic
\subsection{Unresolved DNS Query Traffic}

The DNS traffic data is aggregated daily over three years. The DNS traffic does not exhibit any obvious seasonal components and represent a wide range of values. In one day, a network could receive as few as 46 DNS queries to as many as almost three-hundred million. Hence, allocating a static amount of network resources that is sufficient to service the maximum observed number of unresolved DNS queries is wasteful. This is why, it is essential to predict these peak volumes and then allocate a sufficient number of resources only when it is necessary. Linear Interpolation is applied to fill in any gaps within the data set as described in Table~\ref{tab:data_stats} and in Fig.~\ref{fig:dns_linear_plot}.

Even though the Unresolved DNS queries data does not exhibit any obvious seasonal patterns with the data, PPFD did improve the performance for peak values demonstrated by ``Peak RMSE'' and ``Peak RWSE'' in Table~\ref{tab:dns_results_linear} for both ANN and ARIMA. Moreover, looking at Fig.~\ref{fig:dns_ann_vs_fourier_ann}, there are three prominent peaks, encapsulated by the red boxes, in the DNS query data set that are now being over predicted or predicted closer to the expected series rather than under predicted by the baseline. Due to the non-obvious seasonal patterns, PPFD with larger values of $c$ generally increases the number of peaks that can be over predicted rather than under predicted. This demonstrates that the proposed PPFD is able to help predict both peak volumes and general volumes in real tasks, even when the seasonality of the data is not that obvious.

%%%%% Subsection: Electricity Consumption
\subsection{Electricity Consumption}

Lastly, we include an experiment on a real data with more obvious seasonality to test the efficacy of the proposed framework, which is the electricity consumption aggregated hourly. This data set contains large segments of missing data points that are set to 0. There is a gap of 119 continuous missing data points starting on the August $17^{th}$, 2010 at 10pm with another gap of 87 data points after that. Therefore, all data after August $17^{th}$, 2010 at 10pm is thus removed, and then Linear Interpolation is applied as shown in Table~\ref{tab:data_stats} and Fig.~\ref{fig:electricity_linear_plot}. Table~\ref{tab:hourly_power_linear} provides similar findings, which further demonstrates that our proposed method, PPFD, is beneficial for both peak prediction and general prediction.
%%%%%%%%%%%%%%%%%%%%%%%%%%%%%%%%%%%%%%%%%%%%%%%%%%%%%%%%%%%%%%%%%%%%%%%%%%%%%
%%%%% Section: Conclusion
\section{Conclusion}

For entities providing services such as DNS, DHCP, IP address management, it is crucial to anticipate when these services will receive their peak volumes of requests. In this work, we have found that time-series decomposition, particularly PPFD, has improved the performance of general prediction models that have been applied to network traffic in forecasting the peak values and in some circumstances improved the performance for general prediction of all values. In future work, we want to study the effects that missing data has on Fourier Forecasting to assess the robustness of this procedure against missing data. In addition to this investigation, we want to add to this work by testing the effects of our proposed framework on more complex Neural Networks such as RNNs and attention networks.

\section{Acknowledgement}
Stewart and Hu's research is supported in part by NSF (IIS-2104270). All opinions, findings, conclusions and recommendations in this paper are those of the author and do not necessarily reflect the views of the funding agencies.

\bibliographystyle{IEEEtran}
\bibliography{bibliography}

\end{document}